\documentclass[twocolumn,aps,prb,floatfix]{revtex4}

\usepackage{graphics}

\begin{document}

\title{One-band tight-binding model parametrization of the high-$T_c$
cuprates, including the effect of $k_z$-dispersion}

\author{R.S. Markiewicz$^1$, S. Sahrakorpi$^1$, M. Lindroos$^{1,2}$,
Hsin Lin$^1$, and A. Bansil$^1$}

\affiliation{
1: Physics Department, Northeastern University, Boston MA
02115, USA\\
2: Institute of Physics, Tampere University of Technology, P.O. Box
692, 33101 Tampere, Finland}

\date{\today}

\begin{abstract}

   We discuss the effects of interlayer hopping and the resulting
   $k_z$-dispersion in the cuprates within the framework of the
   one-band tight binding (TB) model Hamiltonian. Specific forms of
   the dispersion relations in terms of the in-plane hopping
   parameters $t$, $t'$, $t''$ and $t'''$ and the effective interlayer
   hopping $t_z$ in La$_{2-x}$Sr$_x$CuO$_4$ (LSCO) and
   Nd$_{2-x}$Ce$_x$CuO$_4$ (NCCO) and the added intracell hopping
   $t_{bi}$ between the CuO$_2$ bilayers in
   Bi$_2$Sr$_2$CaCu$_2$O$_{8}$ (Bi2212) are presented. The values of
   the `bare' parameters are obtained via fits with the first
   principles LDA-based band structures in LSCO, NCCO and Bi2212. The
   corresponding `dressed' parameter sets which account for
   correlation effects beyond the LDA are derived by fitting
   experimental FS maps and dispersions near the Fermi energy in
   optimally doped and overdoped systems. The interlayer couplings
   $t_z$ and $t_{bi}$ are found generally to be a substantial fraction
   of the in-plane hopping $t$, although the value of $t_z$ in NCCO is
   anomalously small, reflecting absence of apical O atoms in the
   crystal structure. Our results provide some insight into the issues
   of the determination of doping from experimental FS maps in Bi2212,
   the role of intercell coupling in $c$-axis transport, and the
   possible correlations between the doping dependencies of the
   binding energies of the Van Hove singularities (VHSs) and various
   prominent features observed in the angle-resolved photoemission
   (ARPES) and tunneling spectra of the cuprates.

\end{abstract}

\maketitle

\section{Introduction}

A good deal of the existing literature on the cuprates invokes the
one-band tight binding (TB) model Hamiltonian based essentially on the
properties of a single CuO$_2$-layer. Such a two-dimensional (2D)
treatment ignores intercell coupling in a single layer material such
as La$_{2-x}$Sr$_x$CuO$_4$ (LSCO) or Nd$_{2-x}$Ce$_x$CuO$_4$ (NCCO)
and the additional intracell couplings in multi-layer materials like
Bi$_2$Sr$_2 $CaCu$_2$O$_{8}$ (Bi2212). The interlayer hopping however
is responsible for inducing finite dispersion of energy bands with
$k_z$ and for splitting the two CuO$_2$ bands in Bi2212 into bonding
and antibonding combinations whose existence has become widely
accepted in the high-$T_c$ community in the last few
years\cite{Bogdanov01,chuang01,asensio03}. Moreover, it has been shown
recently that the angle-resolved photoemission (ARPES) spectra from a
quasi 2D material differ fundamentally from the three-dimensional (3D)
case due to the loss of $k_z$ selectivity\cite{Matti1}. The ARPES
peaks as a result display an irreducible width which does not have its
origin in any scattering mechanism. Such $k_z$-dispersion-induced
linewidths possess characteristic $k_{\parallel}$ dependencies and
have been shown to be quite significant in LSCO\cite{Seppo1} and
Bi2212\cite{Matti1}, complicating the interpretation of spectroscopic
data.

It is clear with this backdrop that it is important to incorporate
interlayer couplings and the associated $k_z$-dispersion effects in
modeling the cuprates. This article represents the first such
comprehensive attempt by considering the examples of the single layer
LSCO and NCCO and the double layer Bi2212 systems. For this purpose,
the framework of the standard dispersion relation, $E(k)=
E(k_{\parallel})$, is generalized to include an additional term
$E_z(k_{\parallel},k_z)$. Our focus is on the form of $E_z$ in LSCO,
NCCO and Bi2212 in terms of the effective intercell, interlayer
hopping parameter $t_z$ and in the case of Bi2212 the added intracell
hopping $t_{bi}$, although forms in LSCO and NCCO involving additional
intercell hopping terms are also presented. The values of various TB
parameters are obtained via fits to the first principles local density
approximation (LDA) based band structures as well as to the
experimental Fermi surfaces (FSs) and dispersions near the Fermi
energy ($E_F$) in the optimally doped and overdoped systems. In
particular, we derive estimates of the interlayer hopping parameters
in terms of the bilayer splitting and the characteristic
$k_{\parallel}$-dependent broadenings of the FS features observed in
FS maps obtained via ARPES experiments. Similarities and differences
between the in-plane and out-of-plane hopping parameters so determined
in LSCO, NCCO and Bi2212 are delineated.  These results are used to
gain some insight into a number of issues of current interest such as
the determination of doping from the experimental FS maps in Bi2212,
the role of intercell coupling in $c$-axis transport, and the possible
correlations between the doping dependencies of the binding energies
of the Van Hove singularities (VHSs) and various prominent features
observed in ARPES and tunneling spectra of the cuprates.

It should be noted that our analysis implicitly assumes the presence
of large FSs and an LDA-type electronic spectrum, albeit with some
renormalization, and it is limited therefore generally to the
description of the optimally doped and overdoped cuprates. The
situation with underdoping becomes more complicated, especially in
LSCO and NCCO, as a lower Hubbard band with an increasing spectral
weight develops, which is not accounted for by the LDA or the present
TB model fits.  However, the hopping parameters of the sort obtained
here based on the ``uncorrelated" LDA-type spectrum form the
traditional starting point for investigating the effects of strong
correlations and are thus of considerable value in exploring the
physics of the cuprates more generally.

An outline of this article is as follows. Section~II presents relevant
forms of energy band dispersions within the framework of the one band
model as applied to LSCO, NCCO and Bi2212. Section~III briefly
describes the procedure used to obtain fitted TB
parameters. Section~IV considers the `bare' parameters which fit the
LDA-based band structures near the Fermi energy and the quality of the
resulting fits. Section~V addresses modifications needed to fit
experimental FSs and dispersions and the associated renormalized or
`dressed' parameters. Section~VI summarizes various bare and dressed
parameter sets and comments on other such datasets available in the
literature. Section~VII briefly discusses a few illustrative examples
of applications of our results. Finally, Section~VIII presents a
summary and makes a few concluding remarks.

\section{Model Dispersions}

\begin{figure}[t]
    \resizebox{7.5cm}{!}{\includegraphics{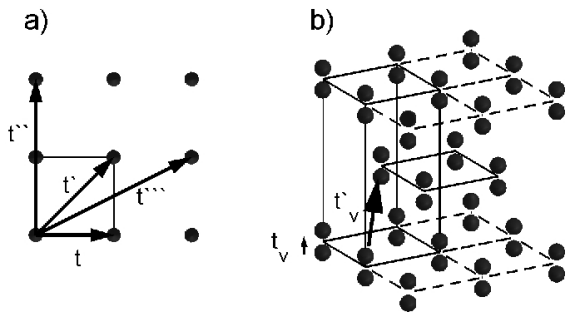}}
\caption{
   Schematic crystal structure of cuprates, showing Cu positions only,
   to illustrate various hopping processes. (a) In-plane hopping
   parameters $t$, $t'$, $t''$ and $t'''$ between different Cu atoms
   in the plane. (b) Simplified out-of plane hopping parameters along
   the $c$-axis.  $t_v$ illustrates hopping within a bilayer in Bi2212
   and $t_v'$ across the unit cells. The actual hopping process is
   more complicated as discussed in the text. Note that the stacking
   of CuO$_2$ layers across adjacent unit cells is staggered,
   i.e. shifted by $(a/2,a/2)$.}
\label{fig:1}
\end{figure}
In developing model dispersions for various compounds within a one
band (single orbital on Cu) TB scheme, we begin by recalling that in
general the three-dimensional energy bands for the TB Hamiltonian can
be expressed as (assuming a lattice without a basis):
\begin{equation}
   E({\bf k})=\sum_{\bf l} exp(i{\bf k\cdot l})t_{\bf l} \>,
\label{eq:0b}
\end{equation}
where $\bf {l}$ denotes a lattice vector and $t_{\bf l}$ is the
hopping parameter associated with the $l^{th}$ site. Since our focus
is on one-band models, we highlight the arrangement of the Cu atoms
within the CuO$_2$ planes as well as between different CuO$_2$ planes
in Fig.~1, where various intraplane and interplane hoppings are
identified.

In order to emphasize the effect of $k_z$-dispersion, we write
\begin{equation}
   E({\bf k})=E_{\parallel}({\bf k}_{\parallel})
   + E_z({\bf k}_{\parallel},k_z) \>.
\label{eq:0c}
\end{equation}
Here, ${\bf k}_{\parallel}$ and $k_z$ respectively denote the in-plane
and out-of-plane components of $\bf k$. $E_{\parallel}$ is the model
dispersion associated with a single CuO$_2$ plane if the effects of
$k_z$-dispersion are totally neglected. The $E_z$ term accounts for
the effect of coupling between different CuO$_2$ planes, and possesses
a very different form in a single layer cuprate such as LSCO or NCCO
than when a CuO$_2$ bilayer is present as in Bi2212. We now discuss
the two terms on the right hand side of Eq.~\ref{eq:0c} in turn.

\subsection{In-plane term $E_{\parallel}$}

$E_{\parallel}$ is straightforwardly expressed via the hoppings $t$,
$t'$, $t''$ and $t'''$ of Fig.~1(a) as
\begin{eqnarray}
   E_{\parallel}({\bf
   k}_{\parallel})=-2t[c_x(a)+c_y(a)]-4t'c_x(a)c_y(a) \nonumber \\
   -2t''[c_x(2a)+c_y(2a)] \nonumber \\
   -2t'''[c_x(2a)c_y(a)+c_y(2a)c_x(a)] \> ,
\label{eq:2}
\end{eqnarray} 

where
\begin{equation}
   c_i(\alpha a)\equiv cos(\alpha k_ia) \>,
\label{eq:0d}
\end{equation}
and $\alpha$ is an integer (or half-integer).  The shorthand notation
of Eq.~\ref{eq:0d} is convenient as well as transparent since factors
of $cos(\alpha k_ia)$ occur frequently in describing hoppings between
lattice sites separated by different distances $\alpha a$.  Only
$\alpha$ values of 1 and 2 occur in Eq.~\ref{eq:0d} because hopping up
to a maximum of two lattice sites in the $x$ and/or $y$ direction is
included as seen from Fig.~1(a).

\subsection {Interlayer term $E_z$}

The form of the interlayer component $E_z$ depends sensitively on the
stacking of the layers. Consider an isolated bilayer first. From
Fig.~1(b), the simplest vertical hop is straight up via $t_v$. Such a
term however in Eq.~\ref{eq:0c} will be independent of
$k_{\parallel}$.  Realistic band computations show that this is not
the case and that the bilayer split bands in Bi2212 display
considerable dispersion with $k_{\parallel}$. In particular,
Ref.~\onlinecite{PWA} adduced
\begin{equation}
   E_{bi}=-2t_{bi}(c_x(a)-c_y(a))^2 \>.
\label{eq:2c}
\end{equation}
\noindent
It should be emphasized that $t_{bi}$ here is an effective parameter
for hopping within a single bilayer. By expanding the right side of
Eq.~5 as
\begin{equation}
   (c_x(a)-c_y(a))^2=1-2c_x(a)c_y(a)+[c_x(2a)+c_y(2a)]/2 \>,
\label{eq:0e}
\end{equation}
\noindent
it is easily seen that $E_{bi}$ in fact involves the sum of three
different contributions and that its $k_{\parallel}$ dependence is the
result of various in-plane hoppings via the factors of $c_x$ and
$c_y$. [The first term of unity represents the vertical hop, $t_v$,
which is independent of $k_{\parallel}$.] Ref.~\onlinecite{OKA} shows
how form~\ref{eq:0c} for the bilayer splitting arises in the band
structure of Bi2212 through hopping from Cu $d_{x^2-y^2}$ to Cu $4s$
via the planar oxygen orbitals, with Cu $4s$ states playing a dominant
role.

\subsubsection{LSCO}

In LSCO, the dominant intercell hopping has a form similar to that of
Eq.~\ref{eq:2c}, modified by the body-centered tetragonal structure,
which leads to an offset of successive CuO$_2$ planes by half a unit
cell in the diagonal in-plane direction as seen in
Fig.~\ref{fig:1}(b), resulting in
\begin{equation}
   E_z=-2t_zc_z(c/2)(c_x(a)-c_y(a))^2S_{xy} \> ,
\label{eq:2a}
\end{equation}
where $c$ denotes the lattice constant along the $z$-axis, and
\begin{equation}
   S_{xy} = c_x(a/2)c_y(a/2)
\label{eq:2aa}
\end{equation}
accounts for the presence of the aforementioned layer
offset\cite{OKA2}.  The term $c_z$ arises because we have an infinite
number of stacked layers, whereas the expression of Eq.~\ref{eq:2c}
refers to the case of an isolated pair of layers. Note that $S_{xy}$
vanishes for in-plane momenta along $(\pi ,0) \rightarrow (\pi ,\pi)$,
leading to a lack of $k_z$-dispersion along this high symmetry line,
which is consistent with the observed dispersion in LSCO.

\subsubsection{NCCO}

In NCCO, the $t_z$ term similar to Eq.~\ref{eq:2a} is anomalously
small, due presumably to the absence of the apical oxygens.
Therefore, a number of other competing terms come into play. We
empirically found that the following form approximates the
first-principles band structure results:
\begin{eqnarray}
   E_z=-2c_z(c/2)[t_z(c_x(a)-c_y(a))2+2t'_zs_x(a/2)s_y(a/2) \nonumber\\
   +t''_z(s_x(2a)+s_y(2a))] S_{xy} \> ,
\label{eq:2b}
\end{eqnarray}
where
\begin{equation}
   s_i(\alpha a)\equiv sin(\alpha k_ia) \> ,
\label{eq:0f}
\end{equation}
which is similar to the definition of $c_i$ in Eq.~\ref{eq:0d}.

\subsubsection{Bi2212}

Bi2212 involves two independent interlayer hopping parameters:
$t_{bi}$ which controls the intracell bilayer splitting, and $t_z$,
which involves intercell hopping.  The associated dispersion can be
written as
\begin{equation}
   E_z=- T_z( k_{\parallel},c_z(c/2))[(c_x(a)-c_y(a))^2/4+a_0] \> ,
\label{eq:3}
\end{equation}
where
\begin{equation}
   T_z=\pm\sqrt{t_{bi}^2+A_z^{\prime 2}+2t_{bi}A'_zc_z(c/2)} \> ,
\label{eq:3a}
\end{equation}
plus (minus) sign refers to the bonding (antibonding) solution and
\begin{equation}
   A'_z\equiv 4t_zS_{xy} \> .
\label{eq:3b}
\end{equation}
Note that the $(c_x(a)-c_y(a))^2$ term leads to absence of
$k_z$-dispersion along the $\Gamma \rightarrow (\pi ,\pi )$ line, and
particularly at $\Gamma$. The introduction of the additional,
`vertical' hopping parameter, $a_0$ allows for the presence of a
splitting at $\Gamma$.

\section{Fitting Procedure}

We first fit the first principles band sructures in various compounds
with the TB forms of Eqs.~\ref{eq:2}-~\ref{eq:3b} to obtain what may
be referred to as the `bare' hopping parameters. For this purpose,
self-consistent band structure calculations were carried out within
the LDA using the well-established Green's function
methodology\cite{bansil99_1}. The fitting process starts by
considering the in-plane dispersion and can be illustrated with the
example of the monolayer system. Specifically, the in-plane dispersion
is analyzed in terms of Eq.~\ref{eq:2} by fixing $k_z=\pi /c$ and
using the LDA energies at a set of high symmetry points. The zero of
energy is set for convenience at the energy of the VHS given by the
$(\pi,0)$ point, i.e.,
\begin{equation}
   E_V\equiv E{(\pi ,0)}=4(t'-t'') \> .
\label{eq:3c}
\end{equation}
The energies at four other symmetry points from Eq.~\ref{eq:2} then
are (with respect to $E_V$)
\begin{equation}
   E_1= E(\Gamma )=-4(t+2t''+t''') \> ,
\label{eq:11a}
\end{equation}
\begin{equation}
   E_2=E{(\pi ,\pi )}=4(t-2t''+t''') \> ,
\label{eq:11b}
\end{equation}
\begin{equation}
   E_3=E{(\pi /2,\pi /2)}=4(2t''-t') \> .
\label{eq:11c}
\end{equation}
\begin{equation}
   E_4=E{(\pi ,\pi /2)}=2(t-t''')+4(t''-t') \> .
\label{eq:11d}
\end{equation}

Eqs.~\ref{eq:11a}-~\ref{eq:11d} can be inverted straightforwardly to
obtain the four unknowns, $t,t',t''$ and $t'''$ in terms of the four
LDA energies $E_1-E_4$. In terms of the bandwidth
\begin{equation}
   W=E_2-E_1=8(t+t''')
\label{eq:11i}
\end{equation}
we have
\begin{equation}
   t=(E_4+E_V+W/4)/4 \> ,
\label{eq:11e}
\end{equation}
\begin{equation}
   t'=(W/2- E_2)/8 \> ,
\label{eq:11h}
\end{equation}
\begin{equation}
   t''=t'/2+E_3/8 \> ,
\label{eq:11g}
\end{equation}
and
\begin{equation}
   t'''=W/8-t \> .
\label{eq:11f}
\end{equation}
The resulting dispersions do not completely agree with the LDA,
indicating that more distant hops are significant. Instead of
including these, we have adjusted the parameters slightly to optimize
the overall fit. In Bi2212 we apply the above procedure to the average
of the two bilayer split bands.  Once the in-plane parameters are
determined, the out-of-plane ($k_z$) dispersion is fitted along
similar lines using Eqs.~\ref{eq:2a}-~\ref{eq:3b} to obtain the
interlayer hopping parameters.

\begin{figure}[t]
    \resizebox{7.5cm}{!}{\includegraphics{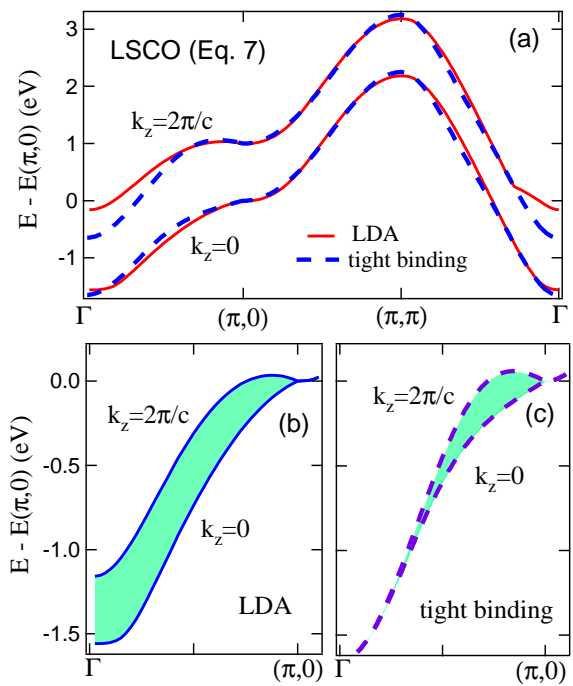}}
\caption{
   Comparison of first principles energy bands in LSCO (solid lines)
   with the corresponding tight binding fit based on
   Eq~\protect\ref{eq:2a} (dashed lines). (a) Dispersion as a function
   of in-plane momentum $k_{\parallel}$ along several high symmetry
   lines for $k_z=0$ and $k_z=2\pi/c$. Upper set of bands is shifted
   by 1 eV with respect to the lower set for clarity.  $E(\pi,0)$,
   which is the energy of the VHS, defines the energy zero. (b) Effect
   of $k_z$-dispersion as a function of $k_{\parallel}$ for LDA bands.
   Shaded region is bounded by the bands for $k_z=0$ and
   $k_z=2\pi/c$. (c) Same as (b), except that this figure refers to
   the corresponding results for the fitted bands.}
\label{fig:20A}
\end{figure}
It should be noted that the bottom of the valence band in the cuprates
usually overlaps with other bands derived from planar and apical
O-atoms.  This makes it difficult to locate the position of the band
bottom $E_1$ in LDA band structures. For this reason, we have used an
average value of LDA levels in the vicinity of the band bottom to
define $E_1$ used in Eq.~\ref{eq:11a}.

The measured FSs and band dispersions in the cuprates in general do
not agree with the LDA results, indicating the presence of electron
correlation effects beyong the LDA. The bare TB parameters discussed
so far can be modified or `dressed' to fit the experimental data in
optimally doped and overdoped systems which exhibit large hole-like
FSs. The details of the fitting procedure vary somewhat for the
different compounds considered and are discussed below. While the
values of the dressed and bare parameters differ significantly, the
main effect appears to be a renormalization (decrease) of the {\it
bandwidth} by a multiplicative factor $Z$, with little effect on the
{\it shape} of the FS.

\section{Bare Parameters}

In this section we discuss the bare parameters and the extent to which
they reproduce the LDA band structures and FSs.

\subsection{LSCO}

The fits in LSCO based on Eqs.~\ref{eq:2} and~\ref{eq:2a} yield:
$t=0.43$ eV, $t'/t=-0.09$, $t''/t=0.07$, $t'''/t=0.16$ and
$t_z/t=0.12$. The interlayer hopping $t_z$ is a significant fraction
of the in-plane nearest neighbor (NN) hopping $t$ and the size of
$t_z$ is comparable to the values of in-plane second and higher
neighbor terms.  Fig.~\ref{fig:20A} considers the quality of the TB
fit. The LDA and TB bands as a function of $k_{\parallel}$ at $k_z=0$
as well as at $k_z=2\pi/c$ are seen from Fig.~\ref{fig:20A}(a) to be
in good accord near the region of the VHS (defined as energy zero),
which is the region of greatest interest in delineating near-FS
physics.  The $k_{\parallel}$ dependence of the $k_z$ dispersion and
how it differs between the TB and the LDA is better seen by comparing
(b) and~(c), where the shaded regions give the energy range over which
the band wanders as a function of $k_z$.

\begin{figure}[t]
    \resizebox{7.5cm}{!}{\includegraphics{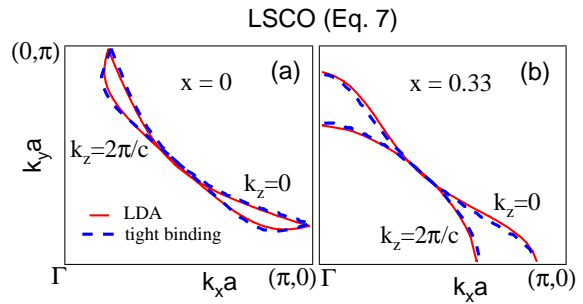}}
\caption{
   FS maps as a function of $k_{\parallel}$ for different $k_z$ values
   in LSCO are compared for the LDA (solid lines) and the
   corresponding tight binding fit (dashed lines) using
   Eq.~\protect\ref{eq:2a}.  Results for doping levels of $x=0$ (a)
   and $x=0.33$ (b) are shown.}
\label{fig:2A}
\end{figure}
Fig.~\ref{fig:2A} considers FS maps. Given the good fit to the LDA
bands near the VHS noted above, the agreement between the TB and LDA
FSs is to be expected.  The TB fit correctly reproduces the cuts
through the FS at $k_z=0$ and $k_z=2\pi/c$ as well as changes in the
shape of the FS as one goes from the undoped case in (a) to the highly
doped case of $x=0.33$ in (b).

\begin{figure}[t]
    \resizebox{7.5cm}{!}{\includegraphics{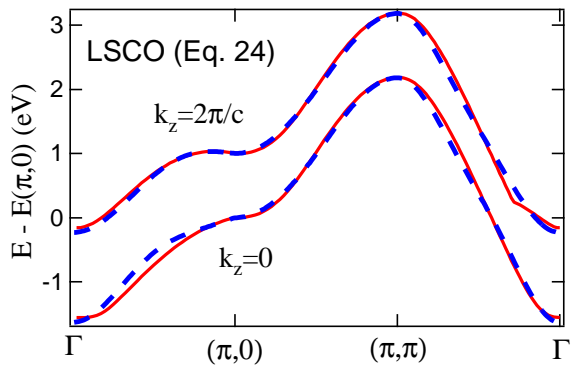}}
\caption{
   Same as Fig.~\protect\ref{fig:20A}(a), except that the TB fit
   (dashed lines) here is based on Eq.~\protect\ref{eq:2ac} which
   better reproduces the LDA result (solid lines) near $\Gamma$.}
\label{fig:20}
\end{figure}
Notably, the TB fit in Fig.~\ref{fig:20A} deviates significantly from
the LDA bands around $\Gamma$ for $k_z=2\pi/c$. This is not surprising
since, as already pointed out, the band bottom overlaps with other
Cu-derived bands. Therefore, a proper solution to this problem
requires the use of a multi-band TB model. Nevertheless, we have
empirically found that the following form within the single band model
gives a reasonably good fit.
\begin{equation}
   E_z=-2(t_zc_z(1/2)+t_z'c_z^2(1/2))[(c_x(1)-c_y(1))2+a_0
   S_{xy}^2]S_{xy} \> ,
\label{eq:2ac}
\end{equation}
where an additional hopping parameter $t_z'$ has been introduced
together with the constant $a_0$; other quantities here have been
defined earlier.  The fit based on Eqs.~\ref{eq:2} and~\ref{eq:2ac}
gives: $t=0.4$ eV, $t'/t=-0.125$, $t''/t=0.05$, $t'''/t=0.125$,
$t_z/t=0.125$, $t_z'/t=0.05$, and $a_0=0.083$. By comparing these
values with those given above based on Eqs.~\ref{eq:2}
and~\ref{eq:2a}, we see that the in-plane parameters are quite similar
in the two sets of fits and are thus quite robust. Fig.~\ref{fig:20}
shows that the fit based on Eq.~\ref{eq:2ac} corrects the discrepency
seen in Fig.~\ref{fig:20A} near the $\Gamma-$point.

Although Eq.~\ref{eq:2ac} provides a better fit to the LDA bands, we
should keep in mind that there is no clear evidence for the presence
of an array of bands around a binding energy of $\sim 1$ eV in the
cuprates. It is possible that electron correlation effects lift the
$d_{x^2-y^2}$ band above the complex of Cu- and O-related bands. If
so, Eq.~\ref{eq:2a} may be better suited than Eq.~\ref{eq:2ac} for
describing the physical system.\cite{LSCOms}

\subsection{NCCO}

\begin{figure}[t]
    \resizebox{7.5cm}{!}{\includegraphics{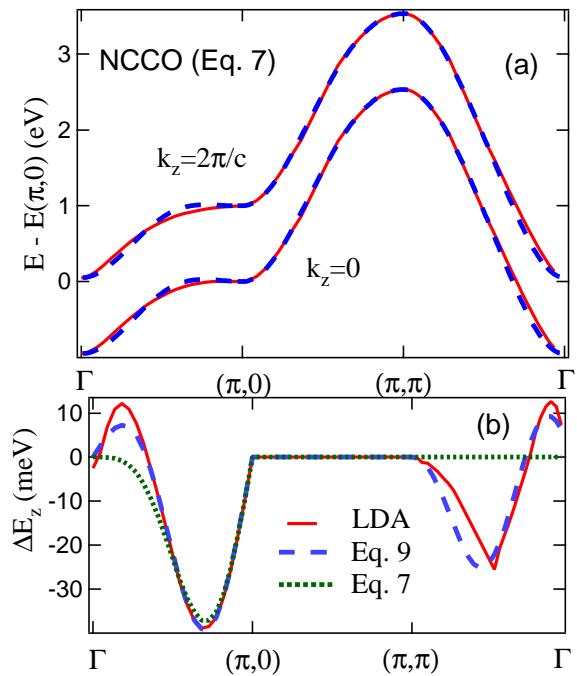}}
\caption{
   (a) Same as Fig.~\protect\ref{fig:20A}(a), except that this figure
   refers to NCCO using Eq.~\protect\ref{eq:2a}. (b) Effect of
   $k_z$-dispersion is shown as a function of $k_{\parallel}$ in the
   form of the spread $E(k_z=0)-E(k_z=2\pi /c)$ for LDA (solid line)
   and fits based on Eqs.~\protect\ref{eq:2a} (dotted line)
   and~\protect\ref{eq:2b} (dashed line).}
\label{fig:8}
\end{figure}
In NCCO, fits were made by using Eq.~\ref{eq:2a} as well as the more
complicated Eq.~\ref{eq:2b} for $E_z$. The resulting in-plane
parameters are: $t=0.42$ eV, $t'/t=-0.24$, $t''/t=0.15$, and
$t'''/t=0.04$. For interlayer parameters, $t_z/t \simeq -0.02$ for
both fits, and the additional parameters of Eq.~\ref{eq:2b} are:
$t_z'/t=-0.025$ and $t_z''/t=0.005$.  The interlayer coupling is much
smaller than in LSCO, reflecting the absence of apical O-atoms, as
already noted.  Figure~\ref{fig:8}(a) shows that the TB fit to the LDA
bands is quite good. The differences between the TB and the LDA bands
are at the meV level and are displayed in the plot of $\Delta
E_z\equiv E(k_z=0)-E(k_z=2\pi /c)$ in Fig.~\ref{fig:8}(b). The simple
form of interlayer hopping in Eq.~\ref{eq:2a} (dotted line in
Fig.~\ref{fig:8}(b)) explains only part of the $k_z$-dispersion, and
the added terms in Eq.~\ref{eq:2b} (dashed line) improve the fit.  The
fitted FSs based on either Eq.~\ref{eq:2a} or Eq.~\ref{eq:2b} are in
very good accord with the LDA and are not shown in the interest of
brevity.

\subsection{Bi2212}

\begin{figure}[t]
    \resizebox{7.5cm}{!}{\includegraphics{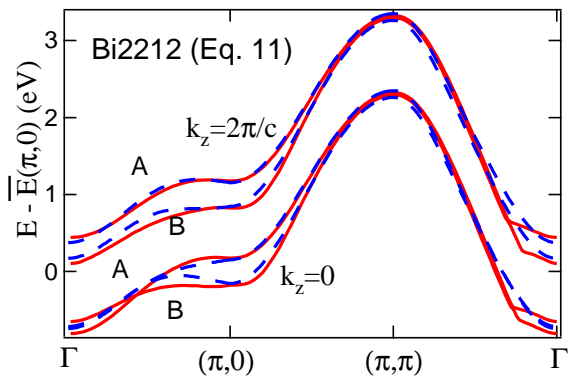}}
\caption{
   Same as Fig.~\protect\ref{fig:20A}(a), except that this figure
   refers to Bi2212 and is fit using Eq.~\protect\ref{eq:3}. Note that
   the bilayer split bonding (B) and anti-bonding (A) bands are shown
   for each $k_z$ value.}
\label{fig:4b}
\end{figure}
The fits on Bi2212 are based on Eq.~\ref{eq:3} and the derived
parameters are $t=0.36$ eV, $t'/t=-0.28$, $t''/t=0.1$, $t'''/t=0.06$,
$t_{bi}/t=0.3$, $t_z/t=0.1$, and $a_0=0.4$. The coupling $t_{bi}$
within a bilayer and the intercell coupling $t_z$ are both quite
substantial as are the in-plane hopping terms beyond the NN
term. Figure~\ref{fig:4b} shows that the TB fit reasonably reproduces
the bilayer split LDA bands despite some discrepancies. The fits for
the FS (not shown) are also quite good.

\begin{figure}[b]
    \resizebox{7.5cm}{!}{\includegraphics{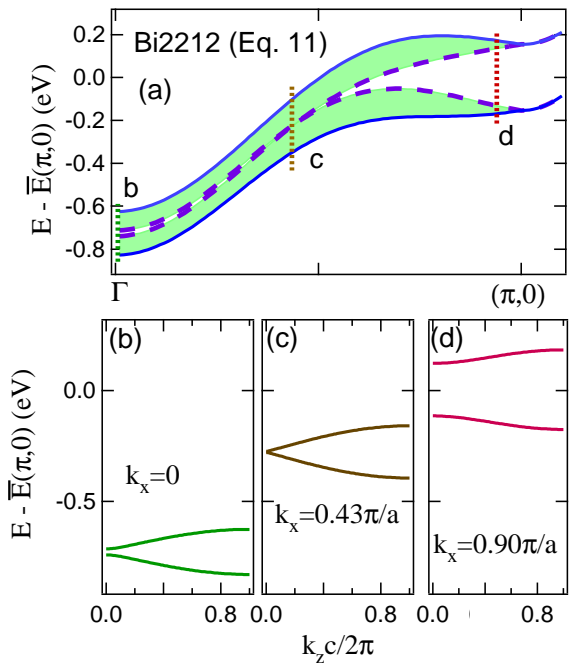}}
\caption{
   (a) Effect of $k_z$-dispersion for the TB fit in Bi2212 showing the
   crossing between the bonding and antibonding bands around
   $k_x=0.43\pi /a$. Shaded areas are bounded by bands for
   $k_{\parallel}$ along the $\Gamma-X$ line for $k_z=0$ (dashed
   lines) and $k_z=2\pi/c$ (solid lines). (b)-(d) Dispersion of bands
   with $k_z$ at three different values of $k_x$ (shown as vertical
   dotted lines in (a)) showing change in the form of $k_z$-dispersion
   through the band crossing point in (a).}
\label{fig:6}
\end{figure}
The lower pair of bilayer split LDA as well as TB bands in
Fig.~\ref{fig:4b} for $k_z=0$ show an interesting level crossing
between the B (bonding) and A (antibonding) bands around ${\bf k} =
{\bf k}_{\parallel}^*=(0,0.4\pi /a,0)$. This level crossing is
associated with an anomalous $k_z$-dispersion, which is highlighted in
Fig.~\ref{fig:6}: the TB bands are drawn on an expanded scale along
the $\Gamma -(\pi ,0)$ line in (a), while the $k_z$-dispersion at
three illustrative $k_{\parallel}$ values is shown in (b)-(d).  The
shape changes from being described approximately by the form
$\sin^2(k_zc/4)$ for $k_\parallel = 0$, to $|\sin(k_zc/4)|$ at
$k^*_\parallel$, and finally back to $\sin^2(k_zc/4)$ for $k_\parallel
a/(\pi) = 1.0$.

\section{Dressed Parameters}

In this section we discuss modified or `dressed' TB parameters needed
for describing the measured FSs and dispersions. As already pointed
out in the Introduction above, our analysis here implicitly assumes
the presence of large LDA-type FSs, and it is thus limited generally
to the optimally doped and overdoped systems.

We emphasize that comparisons between the measured and computed FS
maps only yield information on the relative sizes of the hopping
parameters, e.g. the {\it ratios} $t_i/t$ for various $t_i$'s in terms
of the NN hopping $t$, and that dispersions are additionally needed to
derive the energy scale $t$.  Unfortunately, experimental ARPES data
on dispersions are typically available only over a limited binding
energy range of about 1 eV, leading to inherent uncertainty in any
fitting process. Our procedure for obtaining the dressed parameters
starts by comparing the experimental FSs with the ones computed by
using the bare parameters of the previous section. In this connection,
we compute FS imprints in the $(k_x, k_y)$-plane for $k_z=0$ and
$k_z=2\pi/c$, in order to identify the momentum region over which the
ARPES intensity is allowed due to the effect of $k_z$-dispersion along
the lines of Refs.~\onlinecite{Matti1} and~\onlinecite{Seppo1}.  In
this way, the interlayer coupling parameters can be estimated. All
parameters are then adjusted to obtain a good fit. Fig.~8 gives
insight into these results and the quality of the final FS fits in the
three systems considered. Finally, the measured dispersions are fitted
to determine the average value of the renormalization factor $\bar
Z$. Further relevant specifics in the individual cases of LSCO, NCCO
and Bi2212 are given in the Subsections~V~A-C below. In the interest
of brevity, the final values of all dressed parameters are listed in
Table~II below.

\subsection{LSCO}

\begin{figure}[t]
    \resizebox{7.5cm}{!}{\includegraphics{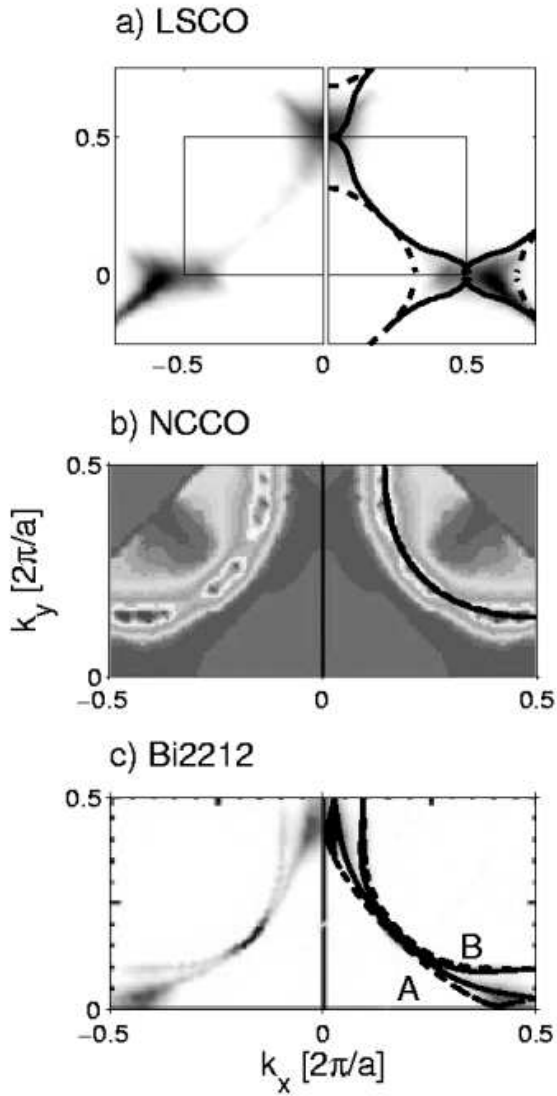}}
\caption{
   Left hand column gives experimental FS maps (i.e. ARPES intensities
   for emission from the Fermi energy) for: (a) An optimally doped
   LSCO sample with $x=0.15$\protect\cite{zhou01,zhouprivate}; (b) An
   optimally doped NCCO sample with $x=0.15$
   \protect\cite{nparm2,expdata}; (c) An overdoped Bi2212
   sample\protect\cite{expdata} (identified as OD70 in
   Ref.~\protect\onlinecite{Bogdanov02}). Right hand side panels give
   the same spectra together with the computed cross sections of the
   FS in the $(k_x, k_y)$-plane for $k_z=0$ (thick solid lines) and
   $k_z=2\pi/c$ (thick dashed lines); see text for the details of the
   dressed tight-binding parameters used. Note, the solid and dashed
   FS contours are well separated for the LSCO and the antibonding FS
   sheet (A) in Bi2212, but this is not the case for NCCO and the
   bonding sheet (B) of the FS in Bi2212.}
\label{fig:5b}
\end{figure}
In LSCO, the experimental FS maps are in good overall accord with the
first principles ARPES computations\cite{Seppo1} over an extended
doping range of $x = 0.06-0.22$, where in the underdoped regime, this
agreement suggests that a remnant of the LDA-type quasiparticles with
a reduced spectral weight continues to persist in the presence of
strong correlations. In this work, we have specifically chosen to fine
tune the bare TB parameters to fit the measured FS
map\cite{zhou01,zhouprivate} at $x=0.15$ using
Eq.~\ref{eq:2a}. Fig.~\ref{fig:5b}(a) is relevant in this regard. The
left hand side shows the measured ARPES intensity from the FS for
reference. The right hand side shows the same intensity, except that
we have overlaid the FS cross sections in the $(k_x, k_y)$-plane for
$k_z=0$ (solid thick lines) and $k_z=2\pi/c$ (dashed thick lines)
obtained by using the final values of the dressed parameters (see
Table~II below).  The parameter ratios are {\it the same} as in the
LDA fit, except for a larger $t'''/t=0.21$.  The experimental
intensity is seen to be reasonably well contained within the
boundaries defined by these FS imprints given by the thick solid and
dashed lines. [The ARPES intensity in this region of allowed
transitions is not expected to be uniform as it gets modulated
considerably in general by the ARPES matrix element.] In particular,
the value of the dressed parameter $t_z$ used reproduces the
characteristic shape of the experimental FS map. Our estimate of the
$t_z$ so obtained should be considered an upper bound since the
effects of experimental resolution and various possible scattering
mechanisms not accounted for in our model will serve to further
broaden the theoretical FS imprint. Finally, we have fitted the
measured dispersion\cite{ino02}, to adduce an average renormalization
factor of $\bar Z=0.64$.  The individual renormalization factors are
$Z(t)$ = $Z(t')$ = $Z(t'')$ = $Z(t_z)$ =0.6, and $Z(t''')=0.8$.

\subsection{NCCO}

In NCCO, the ARPES experiments\cite{nparm,nparm2,nparm3,expdata} seem
to be dominated by the presence of upper and lower Hubbard bands,
which gradually collapse with doping\cite{KLBM}. Near optimal doping
the splitting between the upper and lower Hubbard bands is small and
the resulting FS map resembles the LDA results. Accordingly, we have
used the $x=0.15$ dataset shown in Fig.~8(b) in the fitting process
based on Eq.~7. However, the size of the measured FS is best described
if the doping is taken in the computations to be $x=0.16$, and
therefore, this slightly modified value of doping is used in the
present fits. Also, the observed FS curvature is better fitted by
reversing the sign of the bare $t'''$ (see Tables~I and~II below). Its
effect is to move the FS closer to the magnetic zone boundary given by
the zone diagonal, and thus this sign change may be related to the
presence of some residual anti-ferromagnetic order in the system.
Aside from this change, the fitted in-plane parameter ratios are the
same as the corresponding LDA ratios.

In discussing the bare parameters in Section~IV~B above, we have
already pointed out that the interlayer coupling $t_z$ in NCCO is
quite small. As a result, changes in the cross section of the FS with
$k_z$ are also quite small and for this reason the two FS imprints are
not resolved on the right hand side of Fig.~8(b) (thick solid
line). The situation is in sharp contrast with the case of LSCO in
Fig.~8(a) where the $k_z=0$ and $k_z=2\pi/c$ FS cross-sections are
well separated. These remarks will make it clear that it is difficult
to determine the dressed value of $t_z$ in NCCO with any accuracy via
comparisons of the measured and computed FSs, and therefore, no value
for $t_z$ is given in Table II below.

The TB fit to the FS is seen from Fig.~\ref{fig:5b}(b) to be quite
good overall. The experimentally observed linewidths in ARPES spectra
in NCCO are generally larger than the linewidths that can be
associated with the effect of $k_z$-dispersion. This suggests that
various scattering mechanisms are at play especially around the `hot
spots' where the LDA FS crosses the zone diagonal, allowing for the
opening of pseudogaps which are not accounted for in the LDA or the
present TB fits.  As in the case of LSCO, the final values of the
dressed parameters in NCCO also are more or less uniformly
renormalized with respect to the bare values with a renormalization
$\bar Z=0.55$ for all parameters, except $t'''$ as noted above.

The present TB parameters can be compared to values extracted from
strong coupling calculations. In NCCO, the ARPES data\cite{nparm} were
fit to an antiferromagnetic Slater model\cite{KLBM}.  The hopping
parameters found from the fit are comparable to the bare parameters
with an average renormalization $\bar Z=0.7$ ($Z(t)=0.77$,
$Z(t')=0.84$, $Z(t'')=0.43$).

\subsection{Bi2212}

The right hand side of Figure~\ref{fig:5b}(c) shows the TB fit based
on Eq.~\ref{eq:3} to the experimental normal state FS map of an
overdoped $T_c=70$ K Bi2212 sample, for which the bilayer splitting is
clearly resolved\cite{Bogdanov02,expdata}. For the antibonding FS
sheet, the experimental intensity is seen to fall quite well within
the $k_z=0$ and $k_z=2\pi/c$ slices of the computed FS, especially
near the antinodal region. On the other hand, the bonding FS sheet
displays little width associated with $k_z$-dispersion, which is also
consistent with the character of the measured spectra. The fit in
Fig.~\ref{fig:5b}(c) was used to determine the dressed parameter
ratios, while the overall renormalization factor was found by fitting
the dispersion near $(\pi ,0)$\cite{gromko03,chuang03}.  The
renormalization factors for various parameters are found to be:
$Z(t)=0.35$, $Z(t')=0.36$, $Z(t'')=0.43$, $Z(t''')=0.15$,
$Z(t_{bi})=0.27$, and $Z(t_z) =0.14$. The renormalization of the
hopping parameters is thus seen to be quite substantial with an
average value of $\bar {Z}=0.28 $.  $t_{bi}$ which is responsible for
the bilayer splitting is reduced to about 27\% of its bare
value. Notably, the experimental FS is `pinched' near $(\pi /2,\pi
/2)$, with the nodal point lying closer to the $\Gamma$ point than in
the LDA calculation.

\begin{figure}[t]
    \resizebox{7.5cm}{!}{\includegraphics{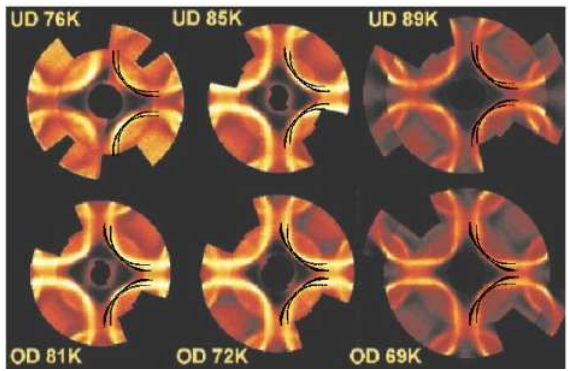}}
\caption{
   Experimental FS maps obtained via ARPES measurements by
   Ref.~\protect\onlinecite{Kord} for six different samples of Bi2212
   covering an extended doping range from underdoped (UD) to overdoped
   (OD) are shown overlaid with the corresponding TB FS maps; see text
   for details of the TB parameters and the Fermi energies used for
   different doping levels. Computed bonding and antibonding FSs are
   shown only in the right portion of each of the six FS maps by black
   lines where the thickness of the lines represents the width of the
   FS induced by the effect of $k_z$-dispersion.}
\label{fig:7}
\end{figure}
Using our dressed TB parameters for Bi2212, we have attempted to model
the doping dependence of the FS of Bi2212 by adjusting the Fermi
energy as a function of doping. [See Section~VII~A below for the
specifics of the Fermi energy values used.] The results are shown in
Fig.~\ref{fig:7}, where the TB FS maps (broadened to reflect the
effect of $k_z$-dispersion) are superposed on the experimental data of
Ref.~\onlinecite{Kord} for a series of dopings from under- to the
overdoped samples. Since the data were taken at room temperature, the
bilayer splitting is not clearly resolved in the experimental
plots. However, the TB fits are seen to reproduce the overall changes
in the shape and size of the FS in Bi2212 with doping reasonably well.

\section{Summary of Tight-binding Parameters}

\begin{table}
\caption{
   Bare TB parameters obtained by fitting the first principles
   LDA-based band sructures as discussed in Section~IV.}
\begin{tabular}{||c||c|c|c||}        
             \hline\hline
Parameter&LSCO&NCCO&Bi2212 \\
     \hline\hline
$t$&430~meV&420~meV&360~meV  \\     \hline
$t'$&-40&-100&-100  \\     \hline
$t''$&30&65&35  \\     \hline
$t'''$&70&15&20  \\     \hline
$t_z$&50&-8&36  \\    \hline
$t_{bi}$ & --- & --- &110 \\     \hline
\end{tabular}
\end{table}
We collect various TB parameter sets in this section for ease of
reference and intercomparisons between the results for LSCO, NCCO and
Bi2212. Table~I lists bare values, while the dressed values are given
in Table~II. The LSCO datasets in Tables~I and~II refer to fits based
on Eq.~\ref{eq:2a}, although a fit using Eq.~\ref{eq:2ac} was also
discussed in Section~IV~A above.  Table~III lists a few other
parameter sets in these compounds available in the literature.
\begin{table}
\caption{
   Dressed TB parameters obtained from fits with the experimental FSs
   and band dispersions as discussed in Section~V. Value of dressed
   $t_z$ in NCCO is undetermined, as discussed in Section~V~B.}
\begin{tabular}{||c||c|c|c||}
             \hline\hline
Parameter&LSCO&NCCO&Bi2212 \\
     \hline\hline
$t$&250~meV&230~meV&126~meV  \\     \hline
$t'$&-25&-55&-36  \\     \hline
$t''$&20&35&15  \\     \hline
$t'''$&55&-30&3  \\     \hline
$t_z$&30&---&5  \\     \hline
$t_{bi}$& --- &---&30 \\     \hline\hline
$\bar Z$&0.64&0.55&0.28 \\ \hline
\end{tabular}
\end{table}

\begin{table*}
\caption{
   Other TB parameter sets determined from LDA band structure
   calculations (columns~1 and~2) or from ARPES experiments
   (columns~3-6).  }
\begin{tabular}{||c||c|c|c|c|c|c||}
             \hline\hline
&LSCO&\multicolumn{4}{c|}{Bi2212}&general \\   \hline
Param.&Ref.~\onlinecite{OKA2}&Ref.~\onlinecite{OKA2}&
Ref.~\onlinecite{Norm}&Ref.~\onlinecite{Norm2}&
Ref.~\onlinecite{Hoog}&Ref.~\onlinecite{Kim}
\\
     \hline\hline
$t$&420~meV&420~meV&150~meV&150~meV&390~meV&350~meV  \\     \hline
$t'$&-73&-110&-40&-24&-82&-120  \\     \hline
$t''$&36&55&13&33&39&80       \\   \hline
$t'''$&--&--&28&13&--&--      \\   \hline
$t^{iv}$&--&--&-13&-24&--&--   \\   \hline
$t_{bi}$&--&--&--&--&90&--    \\   \hline
\end{tabular}
\end{table*}
Some comments concerning the bare parameters in LSCO, NCCO and Bi2212
as well the corresponding dressed values have already been made in
Sections~IV and~V above and need not be repeated. Here we make a few
further observations as follows. In the single layer systems LSCO and
NCCO the dressed values are in excellent agreement with the bare
values, differing only by an overall renormalization $\bar Z$ except
for the most distant neighbor $t'''$.  For Bi2212 bare and dressed
parameters have larger differences (associated with the bilayer
splitting) and the average renormalization factor is smaller,
suggestive of stronger correlation effects.  Given this general
similarity, we restrict our further remarks mainly to the bare
parameter values.

Comparing monolayer LSCO and NCCO, the bare values of the in-plane NN
hopping $t$ are similar ($\sim 0.4$ eV), but other in-plane hoppings,
$t'$ and $t''$ are about twice as large in NCCO compared to LSCO.  For
hole-doped cuprates, an increasing value of $t'$ has been correlated
with increasing superconducting transition
temperature\cite{OKA2,TanZX}, and our results for LSCO and Bi2212 are
consistent with that trend.  The larger $t'$ in NCCO suggests that the
trend does not hold for electron-doped cuprates, but that might be
expected since $E_F$ in electron-doped cuprates is very far from the
VHS.  No clear trends emerge for $t''$ and $t'''$, except that they
are comparable in magnitude to $t'$, and hence should be retained in
detailed analyses.

One may estimate the strength of strong coupling corrections from the
renormalization factors $\bar Z$ via the relation, $\bar
Z=1/(1+\lambda)$, where $\lambda$ is a measure of the combined
strength of electron-electron and electron-phonon coupling.  This
yields $\lambda$ values of 0.6, 0.8 and 2.6 in LSCO, NCCO and Bi2212,
respectively, suggesting much stronger correlations in Bi2212, which
has the highest $T_c$.  These values of $\lambda$ should be used with
caution, since ARPES tends to underestimate dispersion, even in simple
metals\cite{ARooPS}, and a more careful evaluation is needed to
properly account for `kink' physics\cite{Lanzara}.

\begin{figure}[t]
   \resizebox{7.5cm}{!}{\includegraphics{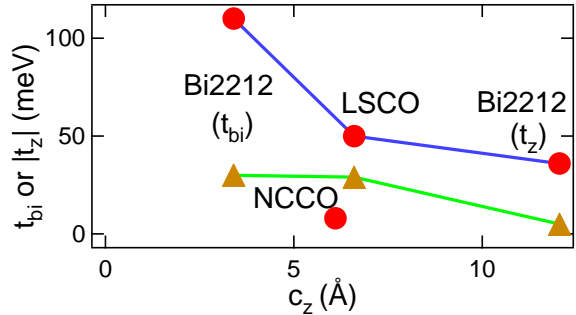}}
\caption{
   Magnitude of the interlayer coupling as a function of the layer
   distance $c_z$ as measured by the values of bare (circles) and
   dressed (triangles) $|t_z|$ and $|t_{bi}|$ in LSCO, NCCO and
   Bi2212. Lines drawn on the data are guides to the eye.}
\label{fig:1b}
\end{figure}
Turning to the nature of interlayer coupling, the values of the bare
and dressed $t_z$ and $t_{bi}$ are plotted in Figure~\ref{fig:1b} as a
function of the separation $c_z$ between CuO$_2$ planes. It is
interesting that this coupling shows a systematic decrease with
increasing separation between the CuO$_2$ layers in Bi2212 and LSCO,
but that NCCO which has a different relative arrangement of Cu and O
atoms in different layers falls outside this pattern. The effect is
more pronounced for the bare parameters (circles) compared to the
dressed ones (triangles).

With regard to other parameter sets for the cuprates in the
literature, note that Table~III is essentially limited to in-plane
couplings since little attention has been paid to the importance of
interlayer coupling in the past. The data of Table~III are generally
similar to the results of Tables~I and~II depending on the type and
details of fit invoked by various authors. More specifically,
Ref.~\onlinecite{OKA2} (first two columns of Table~III) adduces
$t\simeq 0.42$ eV for a large variety of cuprates, and that $t''\sim
-t'/2$; the former result is in agreement with Table~I for LSCO, while
a somewhat smaller value is found for Bi2212.  For Bi2212,
Refs.~\onlinecite{Norm} and~\onlinecite{Norm2} provide two parameter
sets, which involve more distant neighbors than we have included,
although no values for bilayer or other $k_z$ related dispersion are
included.  [The $t^{iv}$ term reflects a contribution $-2t^{iv}c_x(2)
c_y(2)$ to in-plane dispersion.]  These parameter values in Table~III
are seen to be quite close to our dressed values in Table~II for
Bi2212. The parameters found by Ref.~\onlinecite{Hoog} are strikingly
different, although the parameter ratios are more similar.  It may be
that these parameters refer to the total dispersion, while our
evaluation (Table~II) is limited to the region close to the FS.  It is
known that $\bar Z$ is larger in the near-FS regime\cite{Lanzara}.
The last column of Table~III, marked {\em general}, lists the
parameters reported in Ref.~\onlinecite{Kim} as providing a
generalized fit to the experimental ARPES data on a variety of
cuprates and it is difficult to assess their accuracy in the context
of the individual compounds considered in this work.

\section{Illustrative Applications}

\subsection{Doping Dependence of $T_c$ in Bi2212}

The doping level in a given single crystal of Bi2212 can in principle
be determined by measuring the FS and the associated volume and hence
the hole count via an ARPES experiment.\cite{Eis} In practice however
the imprint of the FS in ARPES becomes broadened due to
$k_z$-dispersion and other effects, adding uncertainty to such a
determination of the FS volume in terms of models where the interlayer
coupling is essentially ignored as is the case in much of the existing
literature. Here we attempt to address this issue anew by using our
more complete modeling of the FS in which the effects of
$k_z$-dispersion and bilayer splitting in Bi2212 are explicitly
included.

Our procedure for analyzing the ARPES data in any specific Bi2212
sample is as follows. Using the dressed TB parameters listed in
Table~II, we compute the expected FS imprints for a range of values of
$E_F$, including the effect of broadening of the FS due to
$k_z$-dispersion. [Note that $E_F$ is the only free parameter in the
computation.]  The theoretical FS imprint is then overlaid on the
experimental one to obtain the value of $E_F$ which provides the best
fit to the experimental data.  Once $E_F$ is fixed in this way, the FS
volume and the hole doping is uniquely determined.

\begin{figure}[t]
   \resizebox{7.5cm}{!}{\includegraphics{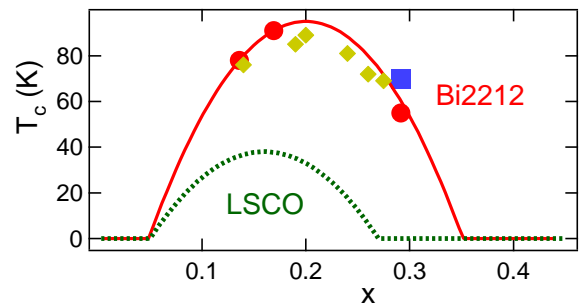}}
\caption{
   $T_c$ vs doping $x$ in Bi2212 as determined by using the Fermi
   energies given in Table~IV and discussed in the text.  Experimental
   data sets are from: Ref.~\protect\onlinecite{Bogdanov02}
   (square),~\protect\onlinecite{chuang03} (solid dots), and
   Ref.~\protect\onlinecite{Kord} (diamonds).  Also shown is the
   $T_c(x)$ dome for LSCO (dotted line), after
   Ref.~\protect\onlinecite{Tal0}.}
\label{fig:9}
\end{figure}

\begin{table}
\caption{
   Estimated values of superconducting transition temperature $T_c$,
   Fermi energy $E_F$ and doping $x$, for various overdoped (OD),
   optimally doped (OP), and underdoped (UD) Bi2212 samples.}
\begin{tabular}{||c|c|c|c|c||}
             \hline\hline
Ref. No.&Doping&$T_c$&$E_F$&x \\
     \hline\hline
\onlinecite{Bogdanov02}&OD&70~K&-160~meV&0.29  \\     \hline
\onlinecite{chuang03}&OD&55&-160&0.29  \\     \hline
\onlinecite{chuang03}&OP&91&-130&0.17  \\     \hline
\onlinecite{chuang03}&UD&78&-120&0.135  \\     \hline
\end{tabular}
\end{table}
Using the preceding approach, we have analyzed the experimental
datasets of Refs.~\onlinecite{Bogdanov02}
and~\onlinecite{chuang03}. The results of the doping values $x$
obtained for different samples are summarized in Table~IV. The best
fit to $E_F$ is also listed in each case. Note that the data cover a
substantial doping range and involve underdoped (UD), optimally doped
(OP) as well as overdoped (OD) samples. The $T_c$ vs $x$ values of
Table~IV are plotted in Fig.~\ref{fig:9} and fall rather nicely on a
parabola.

In their room temperature measurements, Kordyuk, et al.\cite{Kord}
were unable to resolve the bilayer splitting. The doping they
estimated from the area of the single FS was too small, leading them
to hypothesize that they could account for the effect of bilayer
splitting by rigidly shifting the data by a constant $\Delta x =
0.07$. In Fig.~\ref{fig:9}, we show that their data when shifted by
$\Delta x=0.09$ line up reasonably well with other results.  These
shifted values of $x$ in turn give the theoretical FSs plotted in
Fig.~\ref{fig:7} above as broadened black lines.  The residual
differences may indicate subtle differences in the samples\cite{Eis}
or that the actual doping is not exactly described by a rigid shift.
These differences also show up in the theoretical FS maps in
Figs.~\ref{fig:5b}(c) and~\ref{fig:7}.

The solid line in Fig.~\ref{fig:9} gives the parabolic fit through all
the data points and can be expressed as
\begin{equation}
   T_c=T_{c0}[1-({x_0-x\over x_1})^2] \> ,
\label{eq:5}
\end{equation}
\noindent
with $T_{c0}=95$ K being the maximum $T_c$, $x_0=0.2$ is the optimal
doping, and $x_1=0.15$ is the width of the parabola.  This result
should allow the conversion of $T_c$ data to actual doping values in
Bi2212 more generally (as long as it is known whether the sample is
over- or underdoped).

Figure~\ref{fig:9} also compares the superconducting dome of Bi2212,
with that of LSCO\cite{Tal0}, which is given by Eq.~\ref{eq:5} with
$x_0=0.16$, $x_1=0.11$, and $T_{c0}=38$ K.  Both formulas give an
onset of superconductivity at $x=0.05$, while the doping at optimal
$T_c$ in Bi2212 of $x_0=0.2$ is somewhat higher.  Honma, et
al.\cite{HoHo} have recently reported an even higher optimal doping
for Bi2212, $x \sim 0.25$.  The similarity with the Uemura\cite{Uem}
plot is striking, and may give insight into why the Uemura plot peaks
at different dopings for different cuprates\cite{MG}.

\subsection{$c$-Axis Conductivity and Intercell Coupling $t_z$}

Whether the $c$-axis conductivity $\sigma_c$ in the cuprates is
coherent or not is currently a subject of considerable controversy.
However, for a strongly correlated system, one would expect that
increasing the interlayer hopping should increase $\sigma_c$, whether
this conductivity is coherent or not.  For instance, in a fluctuating
stripe model, interlayer conductivity would be sensitive to whether
charged stripes in different layers are in registry, but the actual
hopping will be controlled by $t_z$.  Also, in the interlayer pairing
model, single particle interlayer hopping is frustrated by strong
correlation effects, but the unfrustrated pair hopping still depends
on the bare $t_z$.

\begin{figure}[t]
   \resizebox{7.5cm}{!}{\includegraphics{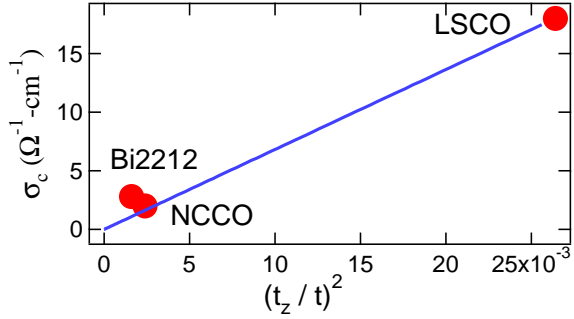}}
\caption{
   $c$-axis conductivity $\sigma_c$ measured just above $T_c$ for
   optimally doped cuprates plotted vs normalized interlayer hopping
   $(t_z/t)^2$. Filled circles give data of
   Ref.~\protect\onlinecite{Dord}, solid line represents expected form
   $\sigma_0 (t_z/t)^2$, with $\sigma_0 =680(\Omega -cm)^{-1}$.}
\label{fig:7b}
\end{figure}

Our purpose here is only to ascertain whether our values of $t_z$ are
consistent with the observed material variations of the $c$-axis
conductivity in the cuprates.  In particular, Ref.~\onlinecite{MKII}
argues that $\sigma_c$ at fixed doping should scale approximately as
$(t_z/t)^2$. Figure~\ref{fig:7b} shows that the plot of
experimental\cite{Dord} $\sigma_c$ vs $(t_z/t)^2$ is nearly
linear\cite{Bibi}, indicating that our $t_z$ values are at least not
obviously inconsistent with experiment in LSCO, NCCO and Bi2212.  Note
that we have taken the $\sigma_c$ values in Fig.~\ref{fig:7b} for the
optimally doped samples where the $c$-axis conductivity is likely to
be more coherent.

\subsection{Effect of $k_z$ Dispersion on Van Hove Singularities}

\begin{figure}[t]
   \resizebox{7.5cm}{!}{\includegraphics{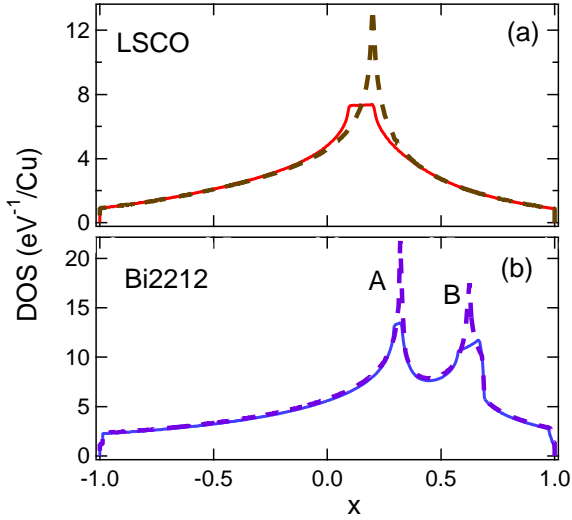}}
\caption{
   Solid lines give TB densities of states (DOSs) for LSCO and Bi2212,
   using dressed parameters of Table II.  Dashed lines give DOSs for
   the limiting 2D case of $t_z$ = 0.}
\label{fig:15}
\end{figure}
Van Hove singularities in the electronic density of states (DOS) have
been implicated for playing an important role in determining various
properties of the cuprates. If the interlayer coupling is neglected,
i.e.  the system is assumed to be strictly 2D, the VHS occurs as a
sharp feature at a single energy in the DOS of a monolayer system and
at two different energies (corresponding to the bilayer split bands)
in a double layer system. Fig.~\ref{fig:15} shows this effect with the
examples of LSCO and Bi2212.  Here the horizontal axis is given in
terms of the doping $x$, which can be obtained straightforwardly for
any given value of the Fermi energy by counting the number of holes in
the band, allowing the energy scale in the DOS to be converted to the
doping scale. The single sharp VHS peak for LSCO in
Fig.~\ref{fig:15}(a) at $x=0.2$ in the 2D case (dashed line) is
replaced by a flat region of high DOS spread over $x=0.09-0.21$ when
the effect of $k_z$-dispersion is included (solid line). A similar
effect is seen in Fig.~\ref{fig:15}(b) for Bi2212, except that here we
have two peaks related to the bonding and antibonding bilayer
bands. The antibonding VHS (A, lower peak around $x=0.32$) displays a
smaller width compared to the bonding VHS (B, around $x=0.62$),
reflecting differences in the extent to which A and B bands are spread
in energy through the coupling $t_{z}$.  By comparing
Figs.~\ref{fig:15}(b) and~\ref{fig:9}, it can be seen that in Bi2212
the optimal $T_c$ corresponds to a doping close to but less that the
doping at which the antibonding VHS crosses the Fermi level.

\begin{figure}[t]
   \resizebox{7.5cm}{!}{\includegraphics{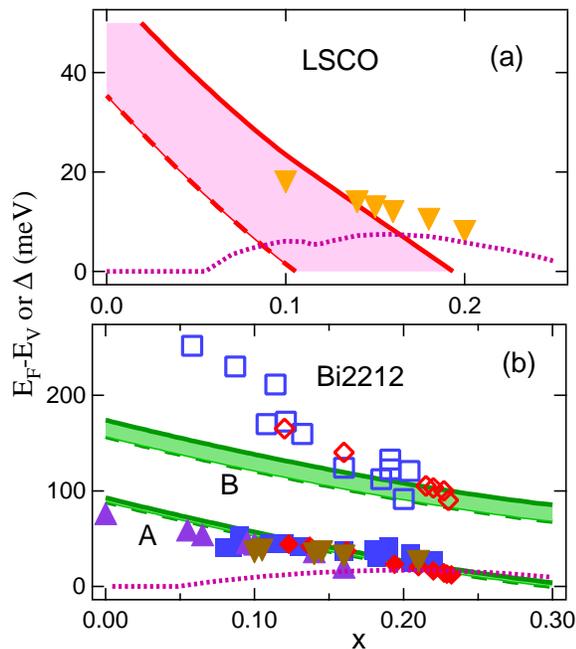}}
\caption{
   Doping dependence of VHS in LSCO (top) and Bi2212 (bottom),
   assuming rigid band filling and using dressed parameters of Table
   II.  The shaded features bounded by solid lines (leading edge) and
   dashed lines (trailing edge) represent the width of the VHS peaks.
   Shown for comparison is the doping dependence of $T_c$, plotted as
   an effective gap, $\Delta =2.05 k_B T_c$ (dotted lines).  Various
   symbols are taken from experiment: In (a)(LSCO) downward triangles
   represent tunneling pseudogaps, Ref.~\protect\onlinecite{Oda}; in
   (b) (Bi2212) tunneling pseudogaps are represented by upward
   triangles (Ref.~\protect\onlinecite{Oda}); ARPES leading edge
   pseudogaps by downward triangles (Ref.~\protect\onlinecite{Marsh})
   and filled squares (Ref.~\protect\onlinecite{Camp}); the ARPES peak
   by filled diamonds (Ref.~\protect\onlinecite{Cuk}); and the ARPES
   hump by open diamonds (Ref.~\protect\onlinecite{Cuk}) and open
   squares (Ref.~\protect\onlinecite{Camp}). }
\label{fig:16}
\end{figure}
The doping dependence of the positions of the VHSs and their possible
correlation with superconducting properties is briefly explored in
Fig.~\ref{fig:16}.  In order to help orient the reader, we consider
Fig.~\ref{fig:16}(a) for LSCO first. The thick solid curve (upper
boundary of the shaded region) gives the position of the leading edge
of the VHS, which is seen from Fig.~\ref{fig:15}(a) to lie at the
Fermi energy for $x=0.19$, and to move to higher binding energies with
decreasing doping. The dashed thick curve similarly gives the binding
energy of the trailing edge of the VHS. Also shown are the gap
parameter $\Delta$, defined via $\Delta =2.05k_BT_c$, for the
superconducting dome (dotted line) and the value of the pseudogap from
tunneling experiments\cite{Oda} (triangles). In the same vein,
Fig.~\ref{fig:16}(b) which refers to Bi2212, gives variations in the
binding energies of the leading and trailing edges of the VHSs for the
bonding and antibonding bands, together with the gap parameter for the
superconducting dome and various features in the tunneling and ARPES
spectra\cite{Marsh,Camp,Cuk} as detailed in the figure caption.

Fig.~\ref{fig:15}(a) shows that the effect of $k_z$-dispersion is
quite substantial on LSCO in that it spreads the VHS over a large
portion of the superconducting dome. In Bi2212 on the other hand the
effect is smaller and the bonding and antibonding VHSs remain
distinct. One of the experimentally observed features in ARPES and
tunneling (the lower binding energy datasets in Fig.~\ref{fig:16}(b))
seems to track the antibonding VHS in Bi2212 over a fairly large
doping range, but this correlation for the higher binding energy
feature is limited to optimal and overdoped regimes. In LSCO, the
tunneling feature shown correlates with the leading edge of the VHS
and is well separated from the behavior of the trailing edge.
However, the connection between the tunneling pseudogap (which is
approximately symmetrical about $E_F$) and the VHS (which is not) is
nontrivial, and a more detailed analysis is considered beyond the
scope of this paper.

\section{ Summary and Conclusions}

We discuss tight-binding parameterization of electronic states near
the Fermi energy in LSCO, NCCO and Bi2212 where the effect of
$k_z$-dispersion is included in a comprehensive manner for the first
time. Our analysis proceeds within the framework of the one-band model
and incorporates available ARPES data and the results of first
principles band structure computations on these cuprates.  The
in-plane coupling is treated in terms of the hoppings $t$, $t'$,
$t''$, and $t'''$ between the nearest and farther out neighbors, while
the coupling between different CuO$_2$ layers is considered mainly
through an effective intercell hopping $t_z$, with an added hopping
$t_{bi}$ to account for the intracell interaction involving the
CuO$_2$ bilayer in Bi2212. In each compound we obtain the specific
tight-binding form for describing the dispersion and the associated
`bare' parameter sets, which fit the 3D band structures based on the
LDA, as well as `dressed' parameters which account for correlation
effects beyond the LDA on the experimental FSs and near-FS dispersions
in optimally doped and overdoped regimes.

In LSCO, the bare interlayer hopping $t_z=0.05$ eV is found to be a
substantial fraction of the in-plane NN hopping $t$ with $t_z/t=0.12$,
and the size of $t_z$ is comparable to the values of in-plane second
and higher neighbor terms. In NCCO, the interlayer coupling is
anomalously small with $t_z/t \simeq -0.02$, reflecting lack of apical
O-atoms in the crystal structure; we delineate how in this case other
smaller terms become important and modify the $k_z$ dependence of the
dispersion. In Bi2212, the bilayer coupling $t_{bi}$ and the
interlayer coupling $t_z$ are both quite substantial, with $t_z/t=0.1$
and $t_{bi}/t=0.3$, as are the in-plane terms beyond the NN term; a
level degeneracy between the bonding and antibonding bands around
${\bf k} =(0,0.4\pi /a,0)$ induces an anomaly in $k_z$
dispersion. Correlation effects beyond the LDA reduce the band
dispersions as expected, so that values of the dressed parameters are
smaller than the corresponding bare values. Renormalization factors
for the hopping parameters vary substantially with average values of
$\bar Z$ = 0.64, 0.55, and 0.28 in LSCO, NCCO and Bi2212,
respectively.

It should be noted that any fitting procedure involving a limited
basis set -- a one band model here -- possesses inherent uncertainty
tied to the binding energy used in obtaining the fit. In the cuprates,
the present framework of the one band model would need to be modified
when one moves to higher binding energies and bands derived from other
Cu $d$- and O $p$ -orbitals come into play. Also, the experimental
values of the interlayer couplings $t_z$ and $t_{bi}$ adduced in this
work should be viewed as upper bounds since these are based on model
computations which only include the broadening effect of
$k_z$-dispersion and do not take into account the effects of various
scattering mechanisms and experimental resolution.

Our modeling of the $k_z$-dispersion effects provides insight into a
number of issues of current interest and we briefly illustrate this
aspect with three examples. (i) We show that the inclusion of
$k_z$-dispersion in modeling FS imprints yields doping levels $x$ for
various published ARPES datasets in Bi2212, which when plotted against
related $T_c$'s, fall remarkably well on a parabola. This
parametrization of the superconducting dome may help the determination
of doping levels of Bi2212 single crystals via ARPES experiments more
generally. (ii) By using the values derived in this work, we show that
the intercell hopping controlled by $t_z$ is consistent with $c$-axis
transport, at least in optimally doped samples, with the $c$-axis
conductivity $\sigma_c$ scaling approximately as $(t_z/t)^2$.  (iii)
Logarithmic Van Hove singularities (VHSs) in the 2D case are spread
over energy when $k_z$-dispersion is included.  We compare the binding
energies of the leading and trailing edges of the VHSs with those of
various prominent features observed in ARPES and tunneling experiments
in LSCO and Bi2212. The antibonding VHS in Bi2212 tracks some
experimental features quite well over a wide doping range, but this is
the case for the bonding VHS in Bi2212 and the leading edge of the VHS
in LSCO mostly in the optimally doped regime. Although specific
implications of these results require further analysis, they suggest
that attention should be given to the $k_z$ induced width of the VHSs
when making comparisons with experimental features.

We conclude that the simple tight-binding parametrization of LSCO,
NCCO and Bi2212 presented in this article constitute a useful starting
point for exploring the effects of $k_z$-dispersion on the electronic
properties of the cuprates.

\begin{acknowledgments}

   This work is supported by the US Department of Energy contract
   DE-AC03-76SF00098 and benefited from the allocation of supercomputer
   time at NERSC and Northeastern University's Advanced Scientific
   Computation Center (ASCC).

\end{acknowledgments}


\begin{thebibliography}{99}
\bibitem{expdata}
%
The experimental data in Fig.~8(b) and~(c) has been extracted from the
figures in Refs.~\onlinecite{nparm2} and~\onlinecite{Bogdanov02},
respectively, and presented here in gray scale.
\bibitem{Bibi}
%
In Bi2212, the relevant parameter is $t_z$ and not $t_{bi}$, where the
latter is related to intracell hopping within a bilayer.
\end{thebibliography}
\end{document}